\documentstyle[prl,aps,epsfig,twocolumn,floats]{revtex}

\newcommand{\M}{{\cal M}}
\newcommand{\eps}{\varepsilon}

\newcommand{\dels}{\delta q^s}
\newcommand{\delv}{\delta q^v}

\newcommand{\ai}{\scriptstyle a_{1}}
\newcommand{\aiNN}{\scriptscriptstyle a_{1}NN}

\newcommand{\bi}{\scriptstyle b_{1}}
\newcommand{\biNN}{\scriptscriptstyle b_{1}NN}
\newcommand{\hi}{\scriptscriptstyle h_{1}}
\newcommand{\hiNN}{\scriptscriptstyle h_{1}NN}
\newcommand{\hit}{h_{1}(1170)}
\newcommand{\hitNN}{\scriptstyle h_{1}(1170)NN}

\newcommand{\gmnn}{g_{\mbox{$\scriptscriptstyle {\M N N}$}}}

\newcommand{\nn}{\nonumber \\}
\newcommand{\be}{\begin{eqnarray}}
\newcommand{\ba}{\begin{array}}
\newcommand{\ea}{\end{array}}
\newcommand{\ee}{\end{eqnarray}}

\newfont{\fib}{cmfi10 at 10pt}

\begin{document}

\draft
\title{\hfill TUHEP-TH01-05\\Flavor-Spin Symmetry
Estimate of the Nucleon Tensor Charge}
\author{Leonard Gamberg\thanks{Electronic mail: gamberg@dept.
physics.upenn.edu}}
\address{Department of Physics and Astronomy, University of
Pennsylvania, David Rittenhouse Labs, 209 South 33rd Street,
Philadelphia, PA 19104\\ and\\
Department of Physics and Astronomy, Tufts University, Medford ,
MA 02155}
\author{Gary R. Goldstein
\thanks{Electronic mail:ggoldste@tufts.edu}}
\address{Department of Physics and Astronomy, Tufts University, Medford ,
MA 02155}
\date{\today}
\maketitle
\begin{abstract}
The axial vector and tensor charge, defined as the first moments of the 
forward nucleon matrix elements of corresponding quark currents, are 
essential for characterizing the spin structure of the nucleon.  
However, the transversity distribution and thus the tensor charge decouple 
at leading twist in deep inelastic scattering, making them hard to measure.  
Additionally, the non-conservation of the tensor charge  makes it difficult 
to predict. There are no definitive theoretical predictions for the tensor 
charge, aside from several model dependent calculations.  We present a new 
approach that exploits the approximate mass degeneracy of the light 
axial vector mesons ($a_1$(1260), $b_1$(1235) and $h_1$(1170)) and 
uses pole dominance to calculate the tensor  charge. The result is simple 
in form. It depends on the decay constants of the axial vector mesons and 
their couplings to the nucleons, along with the 
average transverse momentum of the quarks in the nucleon.
\end{abstract}

The spin composition of the nucleon has been intensely studied
and has produced  important and surprising insights, beginning with the
revelation that the majority of its spin is carried by quark and 
gluonic orbital angular momenta and gluon spin rather than by quark
helicity~\cite{smc,ji_spin}.  Considerable effort has gone into 
understanding, predicting and measuring the 
corresponding transversity distribution of the nucleon
constituents (see~\cite{review} and references therein).  The 
leading twist transversity structure function, $h_1(x)$,
is as fundamental to understanding the nature of the non-perturbative QCD
regime of hadronic physics as is the longitudinal function $g_1(x)$,
which  in principle can be measured in hard scattering processes.
Yet, the transversity distribution is suppressed at leading twist
in deep inelastic lepton scattering since it is chiral odd. The same comparison
applies to the various quark spin dependent  helicity
$\Delta q^{a}(x)$ and
transversity $\delta q^{a}(x)$ (for flavor index $a$) distributions.

In their systematic study of the chiral odd
distributions, Jaffe and Ji~\cite{jaffe91} related the first moment of
the transversity distribution to the flavor contributions
to the nucleon tensor charge:
\be
\int_0^1 \left(\delta q^a(x)-\delta\overline{q}^a(x)\right) dx=\delta q^a .
\ee
Because there must be a helicity flip of the struck quark
in order to probe the transverse spin polarization
of the nucleon,  the transversity distribution
(and thus the tensor charge) decouples at leading twist in deep
inelastic scattering.
No such suppression appears in Drell-Yan scattering where
Ralston and Soper~\cite{ralston79}
first encountered the transversity distribution
entering  the corresponding  transverse spin (both the
beam's and target's spin being transversely polarized to
the incident beam direction) asymmetries. Consequently, the charge is
difficult to measure and its non-conservation~\cite{artru}
makes it difficult to predict.
While bounds placed on the leading twist quark distributions
through positivity constraints suggest
that they  satisfy the inequality of Soffer~\cite{soffer95};
\be
\left|2\delta q^a(x)\right|\le q^a(x) +\Delta q^a(x) ,
\ee
(where $q^a$ denotes the unpolarized quark distribution),
model calculations yield a range of theoretical predictions~\cite{review}.

Like the isoscalar and isovector axial vector charges
defined from the forward nucleon matrix element of the
axial vector current,
the nucleon tensor charge is defined from the corresponding
forward matrix element of the tensor current,
\begin{equation}
\langle P,S_{T}\big|\overline{\psi}
\sigma^{\mu\nu}\gamma_5 \frac{\lambda^a}{2}\psi\big| P,S_{T}\rangle
\hspace{-.05cm}=\hspace{-.05cm}2\delta
q^a(\mu^2)(P^{\mu}S^{\nu}_T\hspace{-.05cm}-\hspace{-.05cm}P^{\nu}S^{\mu}_T),
\label{eq1}
\end{equation}
where ${S_T}^Q \sim \left(\, |+> \pm |->\, \right)$ for the moving
nucleon is the {\it transversity}~\cite{transversity}.
Unlike the axial vector isovector charge, no sum rule
has been written that enables a clear relation between the tensor charge
and a low energy measurable quantity. So, aside from model calculations,
there are no definitive theoretical predictions
of the tensor charge. Among the various approaches, from the QCD sum rule
\cite{jihe,jintang,bely},
which estimates the flavor contributions to the tensor charge by
analyzing the bilocal tensor current,
to lattice calculations~\cite{aoki} and light cone quark
models~\cite{melosh}, there appears to be a range of expectations and a
disagreement concerning the sign of the $d$ quark contribution.

Given the numerous experiments~\cite{phen,star,comp,her}
and proposals~\cite{teslsa,ELFE} for extracting quark transversity
distributions and the nucleon's tensor charge,  there is reason to
expect that in the not too distant future one will be able
to reliably compare the data to the  theoretical estimates of these
quantities.

Here we offer another estimate.
Our motivation stems in part from the result
that the tensor charge does not mix with gluons under QCD
evolution and therefore behaves as a non-singlet
matrix element.
%~\footnote{This is to be contrasted with the axial charges.}
This, in conjunction with the fact that the
tensor current is charge conjugation odd (it does not mix
quark-antiquark excitations of the vacuum, since the latter is charge
conjugation even, nor does it mix with
gluonic operators under evolution, since gluonic operators are even),
suggests that the tensor charge is more amenable to a  valence quark
model analysis. With this in mind  we calculate the
tensor charge by  using axial vector dominance
and an approximate phenomenological mass symmetry among the
light axial vector mesons ($a_1(1260), b_1(1235)$ and $h_1(1170)$).
The $b_1$ and $h_1$ can couple to the quark tensor current in the
nucleon at low energies, and via the symmetry, their coupling to the
leptons is  related to the $a_1$ production in $\tau$ decay.
It is natural to represent this phenomenological mass degeneracy
by $SU(6)_W \otimes O(3)$ spin-flavor symmetry. We utilize
this symmetry in order  to
relate the parameters arising in the expressions
for the tensor charge  to measurable quantities. In the following
we present our results for the isoscalar and isovector and, in turn,
$u$ and $d$ quark contributions to the tensor charge of the
nucleon and  we compare our results  with other model
calculations. Finally we note a remarkable relation between the tensor
charge and the nucleon axial vector coupling constant.

As mentioned above, we  determine  the tensor charge
at a scale where the matrix
element of the tensor current is dominated by the lowest lying axial
vector mesons.~\cite{wein67}
At this scale the matrix element of the tensor
current, Eq.~(\ref{eq1}) is
\be
&&\langle P,S_{T}\Big|\overline{\psi}
\sigma^{\mu\nu}\gamma_5 \frac{\lambda^a}{2}\psi\Big| P,S_{T}\rangle=
 \nn &&\hspace{.5cm}
\sum_{\M} \frac{\langle 0\big|
\overline{\psi}
\sigma^{\mu\nu}\gamma_5 \frac{\lambda^a}{2}
\psi
\big|\M\rangle\langle \M , P,S_{T}| P,S_{T}
\rangle}{M^2_{\M}-k^2} .
\ee
The summation is over those mesons with quantum numbers,
$J^{PC}=1^{+-}$ that  couple to the nucleon via the tensor current;
namely  the charge conjugation odd axial vector mesons -- the isoscalar
$h_1(1170)$ and the isovector $b_1(1235)$.
To analyze this expression in the limit $k^2\rightarrow 0$
we require the vertex functions for the nucleon coupling to the
$h_1$ and $b_1$ meson and the corresponding matrix elements
of the meson decay amplitudes which are related to the meson to vacuum
matrix element via the quark tensor current.
The former yield the nucleon coupling constants
$\gmnn$ defined from the matrix element
\be
&&\langle M P| P\rangle=
\frac{i\gmnn}{2M_N}\overline{u}\left(P,S_{T}\right)
\sigma^{\mu\nu}\gamma_5
u\left(P,S_{T}\right)\varepsilon_\mu P_\nu ,
\ee
where $P_\mu$ is the nucleon  momentum,
and the latter yield the meson decay constant, $f_{\M}$
\be
\langle 0\Big|
\overline{\psi}
\sigma^{\mu\nu}\gamma_5 \frac{\lambda^a}{2}\psi
\Big|\M\rangle=
if^a_{\M}\left(\epsilon_\mu k_\nu-\epsilon_\nu k_\mu\right) ,
\ee
where the $k_\mu$ and $\eps_\nu$ are the meson momentum and
polarization respectively.
For transverse polarized Dirac particles, $S^\mu=(0,S_T)$
we project out the tensor charge using the
constraint on the vector meson, $k\cdot\eps_{\scriptstyle \M}=0$
\be
\delta q^a =
\frac{ f^a_{\M}\  \gmnn \left(S_T\cdot k\right)^2 }{2 M_N
M_{\M}^2}.
\ee
In order to evaluate the tensor charge at the scale dictated
by the axial vector meson dominance  model we must determine the
isoscalar and isovector meson coupling constants.
Taking a  hint from the valence interpretation of the tensor
charge, we  exploit the phenomenological mass
symmetry among the lowest lying axial vector mesons that couple to the
tensor charge; we adopt the spin-flavor symmetry characterized by an
$SU(6) \otimes O(3)$~\cite{sakita,close} multiplet structure.
Thus, the $1^{+-}$ $h_1$ and $b_1$ mesons fall into a
$\left(35\otimes L=1\right)$ multiplet that contains
$J^{PC}=1^{+-},0^{++},1^{++},2^{++}$ states.
This analysis enables us to relate the $a_1$ meson
decay constant measured in $\tau^-\rightarrow a_1^- + \nu_\tau$
decay~\cite{tsai}
\be
f_{\ai} =(0.19\pm 0.03) {\rm GeV^2} ,
\ee
and the  $a_1 N N$ coupling constant
\be
g_{\aiNN}=7.49\pm 1.0 ,
\ee
(as determined from $a_1$ axial vector dominance for longitudinal
 charge as derived by Birkel and Fritzsch~\cite{birkel} but using
$g_A/ g_V= 1.267$~\cite{pdg})
to the meson decay constants, $f_{\bi} $ , $f_{\hi} $ and coupling
constants, $g_{\biNN}$ and $g_{\hiNN}$. We find
%%%%%%%%%%%changes 02jul01
\be
f_{\bi}= \frac{\sqrt{2}}{M_{\bi}}f_{\ai}\, ,\quad
g_{\biNN}=\frac{5}{3 \sqrt{2}} g_{\aiNN}\, ,
\label{eqsym}
\ee
where the $5/ 3$ appears from the $SU(6)$ factor $(1+F/ D)$
and the $\sqrt{2}$ arises from the $L=1$ relation between
the $1^{++}$ and $1^{+-}$ states. Our resulting value
of $f_{\bi}\approx 0.21\pm 0.03$ agrees well with a
sum rule determination of $0.18\pm 0.03$\protect\cite{bely,ball}.
The $h_1$ couplings are related to the $b_1$ couplings via $SU(3)$
and the $SU(6)$ $F/D$ value,
\be
f_{\bi}=\sqrt{3}f_{\hi} \, , \quad
g_{\biNN}=\frac{5}{\sqrt{3}}g_{\hiNN}.
\label{isospin}
\ee
These, in turn, enable us to determine the isovector and isoscalar parts
of the tensor charge,
\be
\delv =\frac{f_{\bi}g_{\biNN}\langle k_{\perp}^2\rangle }{\sqrt{2} M_N
M_{\bi}^2}\, ,\quad
\dels = \frac{ f_{\hi}g_{\hiNN}\langle k_{\perp}^2\rangle }{\sqrt{2} M_N
M_{\hi}^2},
\ee
respectively (where,
$\delv\hspace{-.1cm}=\hspace{-.1cm}(\delta u-\delta d)$, and
$\dels\hspace{-.1cm}=\hspace{-.1cm}(\delta u+\delta d)$).
Transverse momentum appears in these expressions because the tensor
couplings involve helicity flips that carry kinematic factors of
momentum transfer. The intrinsic $k_{\perp}$ of the quarks in the
nucleon is non-zero, as determined from Drell-Yan and heavy vector boson
production processes. Using a Gaussian momentum distribution, and
letting $\langle k_{\perp}^2\rangle$ range from
$\left(0.58 \:{\rm to}\: 1.0\   {\rm GeV}^2\right)$~\cite{ellis}
results in the $u$ and $d$ quark transversity ranging from
\be
\delta u(\mu^2)&=&(0.43 \:{\rm to}\: 0.74)\pm 0.20\, ,
\nn
\delta d(\mu^2)&=&-(0.26 \:{\rm to}\: 0.47)\pm 0.20.
\label{charges}
\ee

These  values for the $u$-quark tensor charge lie slightly lower
than most other estimates while the $d$-quark charge is negative and of a
comparable magnitude. These results, with their uncertainty, are
consistent with the Soffer inequality~\cite{soffer95} applied to the
charges, $q + \Delta q \ge 2|\delta q|$, although near the
equality.  Note that many predictions have the
ratio $\delta d/\delta u$ near $-1/4$ or $(1-\sqrt{3})/(1+\sqrt{3})$,
the value resulting
from an $SU(3)$ degeneracy between the $\pi^0$ and the $\eta(8)$ octet
elements in their coupling to the $u$-quark and the $d$-quark, i.e. the
isoscalar coupling to $u$ and $d$-quarks is $1/\sqrt{3}$ times the
isovector. In our calculation the isoscalar and isovector axial vector
couplings to the nucleon also enter as factors in the expressions for
the charges, with the $D/ F$ ratio being 3/2 in exact $SU(6)$. Loosening the
$SU(6)$ constraint and incorporating mixing of the $h_1(1170)$ with the
$h_1(1380)$ will alter the $u$ to $d$ ratio.
%%%%%%%%%%%% new insert - 02jul01

In relating the $b_1(1235)$ and $h_1(1170)$ couplings in
Eq.~(\ref{isospin}) we
assumed that the latter isoscalar was a pure octet element,
$h_1(8)$. Experimentally, \hskip .25cm the higher mass $h_1(1380)$
was seen in the {}$K+\bar{K}+\pi's$ decay
channel~\cite{pdg,abele} while the $h_1(1170)$ was detected in the
multi-pion
channel~\cite{pdg,ando}. This decay pattern indicates that the higher mass
state is strangeonium and decouples from the lighter quarks -- the well
known mixing pattern of the vector meson nonet elements $\omega$ and
$\phi$. If the $h_1$ states are mixed states of the $SU(3)$ octet $h_1(8)$
and singlet $h_1(1)$ analogously, then
\be
h_1(1170)&=&\sqrt{\frac{2}{3}}h_1(8)-\frac{1}{\sqrt{3}}h_1(1)\, ,
\nn
h_1(1380)&=&\frac{1}{\sqrt{3}}h_1(8)+\sqrt{\frac{2}{3}}h_1(1)\, ,
\ee
from which it follows that
\be
f_{\hit} = f_{\bi}\, ,  \quad{\rm and }\quad   g_{\hitNN} =
\frac{3}{5}g_{\biNN},
\label{mixing}
\ee
with the $h_1(1380)$ not coupling to the nucleon (for
$g_{\hi(1)NN}=\sqrt{2}g_{\hi(8)NN}$). These symmetry relations
alter the results in Eq.~(\ref{charges}) to
\be
\delta u(\mu^2)=&&(0.58 \:{\rm to}\: 1.01)\pm 0.20 \, ,
\nn
\delta d(\mu^2)=-&&(0.11 \:{\rm to}\: 0.20)\pm 0.20.
\label{newcharges}
\ee
The values in Eq.~(\ref{newcharges}) are closer to several other model
calculations including: tensor charges on the lattice~\cite{aoki};
QCD sum rules~\cite{jihe}, the bag model~\cite{jaf,jihe}
and a numerically similar Melosh Transform approach~\cite{melosh};
the chiral quark model with
Goldstone boson effects~\cite{suz};
and quark soliton models~\cite{kim_boch,rein,gamweig}.

The calculation has been carried out at the scale $\mu\approx 1$ GeV,
which is set by the nucleon mass as well as being the mean mass of the
axial vector meson multiplet. The appropriate evolution to higher scales
(wherein the Drell-Yan processes are studied) is determined by the
anomalous dimensions of the tensor charge~\cite{artru}
%via
%\be
%\delta q(Q^2)= \left(\frac{\alpha(Q^2)}{\alpha(\mu^2)}\right)
%^{\frac{4}{33-2n_f}}\delta q(\mu^2).
%\ee
which is straightforward but slowly varying.

It is interesting to observe that the symmetry relations that connect
the $b_1$ couplings to the $a_1$ couplings in Eq.~(\ref{eqsym}) can be
used to relate directly the isovector tensor charge to the axial vector
coupling $g_A$. This is accomplished through the $a_1$ dominance
expression for the isovector longitudinal charges derived by Birkel and
Fritzsch~\cite{birkel},
\be
\Delta u - \Delta d = \frac{g_A}{g_V}=
\frac{\sqrt{2} f_{\ai} g_{\aiNN}}{M_{\ai}^2}.
\label{eqbirkel}
\ee
Hence for $\delv$ we have
\be
\delta u -\delta d =\frac{5}{6}\frac{g_A}{g_V}\frac{ M_{\ai}^2}{
M_{\bi}^2}\frac{\langle k_{\perp}^2\rangle}{ M_N M_{\bi}} \, ,
\label{eqgtga}
\ee
a remarkable relation that appears more fundamental than its
derivation. It is important to realize that this relation can hold at the
scale wherein the couplings were specified, the meson masses, but will be
altered at higher scales (logarithmically) by the different evolution
equations for the $\Delta q$ and $\delta q$ charges. To write an analogous
expression for the isoscalar charges
($\Delta u + \Delta d$) would involve the singlet mixing terms and gluon
contributions, as Ref.~\cite{birkel} considers. However, given that the
tensor charge does not involve gluon contributions (and anomalies), it
may be expected that the relation between the $h_1$ and $b_1$ couplings
in the same $SU(3)$ multiplet will lead to a more direct result
\be
\delta u+\delta d = \frac{3}{5}\frac{M_{\bi}^2}{M_{\hi}^2}\delv\, ,
\ee
for the ideally mixed singlet-octet $h_1(1170)$. These are also
remarkable relations that are quite distinct from other predictions.

In conclusion, our axial vector dominance model with
$SU(6)_W \otimes O(3)$ coupling relations provide simple formulae for the
tensor charges. This simplicity belies the considerable subtlety of the
(non-perturbative) hadronic physics that is summarized in those
formulae. We obtain the same order of magnitude as most other
calculation schemes. These results support the view that the underlying
hadronic physics, while quite difficult to formulate from first
principles, is essentially a $1^{+-}$ meson exchange process. Forthcoming
experiments will begin to test this notion.

\section*{Acknowledgments}
G.R.G. thanks X. Ji for a useful discussion and R. L. Jaffe for bringing
the tensor charge problem to his attention. L. G. thanks H. Reinhardt
and H. Weigel of the Theoretical Physics Institute,
University of T\"{u}bingen for their hospitality where
part of this work was completed.   This work is supported in
part by a grant from the U.S. Department of Energy  \# DE-FG02-92ER40702.

\end{document}